\documentclass[%
 reprint,
 amsmath,amssymb,
 aps,
 pre,
]{revtex4-1}

\usepackage{graphicx}
 \graphicspath{{./}
{./figures/}
}
\usepackage{dcolumn}
\usepackage{bm}
\usepackage{color}
\usepackage{hyperref}

\newcommand{\red}[1]{\textcolor{black}{#1}}

\begin{document}

\preprint{APS/123-QED}


\title{Fifth-order finite-difference scheme for Fokker-Planck equations with drift-admitting jumps}


\author{Yaming Chen}
\email[]{chenym-08@163.com}
\author{Xiaogang Deng}
\affiliation{College of Aerospace Science and Engineering, National University of Defense Technology, Changsha 410073, China}


\date{August 7, 2019}

\begin{abstract}
Recently a useful finite-difference scheme was proposed in [Phys. Rev. E 98, 033302 (2018)] to solve Fokker-Planck equations with drift-admitting jumps. However, while the scheme is fifth order for the case with smooth drifts, it is only second order for the case with discontinuous drifts. To rectify this, we propose in this paper an improved scheme that achieves a fifth-order convergence rate for the case with drift-admitting jumps. Numerical experiments are also employed to verify the validity of the scheme.
\end{abstract}


\maketitle

\section{Introduction}

Piecewise-smooth systems perturbed by noise are used as models of physical and biological systems
\cite{Reimann2002,ChaudhuryMettu2008,GoohpattaderChaudhury2012,Gnoli2013}.
The interplay between noise and discontinuities in such systems has attracted considerable attention recently \cite{Gennes2005}. So far, exact solutions for the propagator (or transition probability distribution) of a few simple piecewise-constant or piecewise-linear stochastic differential equations are known. For instance, the propagator is available in closed analytic form for the case with pure dry (also called solid or Coulomb) friction  \cite{CaugheyDienes1961,Karatzas1984,TouchetteStraetenJust2010}. Other analytical results can also be obtained by using path integrals and weak noise approximations \cite{BauleCohenTouchette2010path,BauleTouchetteCohen2011path,ChenJust2013}. For the corresponding first-passage time problems \cite{ChenJust2014} and functionals \cite{ChenJust2014II,Berezin2018}, some analytical results are available as well. However, there are vast piecewise-smooth stochastic systems that cannot be solved analytically. Therefore, effective numerical methods are necessary for us to understand the underlying dynamics of systems.

In this paper, we are interested in developing numerical methods to calculate the propagator of a simple case that can be modeled by the Langevin equation
\begin{equation}
  \dot{v}(t)=\Phi(v)+\sqrt{2D}\xi(t),
  \label{aa}
\end{equation}
where the overdot denotes the time derivative, $\Phi(v)$ is the drift that may be discontinuous at some points, and $D>0$ represents the strength of the Gaussian white noise $\xi(t)$, characterized by
the zero mean $\langle \xi(t)\rangle=0$ and the correlation $\langle\xi(t)\xi(t^\prime)\rangle=\delta(t-t^\prime)$. Here $\langle\dots\rangle$ stands for the average over
all possible realizations of the noise, and $\delta$
denotes the Dirac delta function.
For the initial condition $v(0)=v_0$, let us denote the propagator of $v$ by $p(v,t|v_0,0)$, which satisfies the following Fokker-Planck equation
\begin{equation}
   \partial_t p=-\partial_v[\Phi(v)p]+D\partial^2_v p
\label{ac}
\end{equation}
with the initial condition  $p(v,0|v_0,0)=\delta(v-v_0)$.

To solve Eq.~(\ref{ac}) with drift-admitting jumps, we need to apply two matching conditions at each jump of the drift, i.e., the continuity of the propagator and the continuity of the probability current (or flux)
\begin{equation}
  f(v,t|v_0,0)=-\Phi(v)p+D\partial_v p. \label{ad}
\end{equation}
Analytically, the solution of the propagator is expected to be nonsmooth at the discontinuous points of the drift. However, while computing the derivative of the current, the algorithm presented in Ref.~\cite{ChenDeng2018PRE} uses stencils across discontinuous points, resulting in an only second-order convergence rate for the case with jumps. To rectify this, we propose in this paper a modification to avoid applying schemes that are derived by using stencils across discontinuities,
elevating the scheme from second order to fifth order for the case with drift-admitting jumps.

The rest of this paper is arranged as follows. In Sec.~\ref{sec_2}, we introduce the used grid and some necessary notations. Then in Sec.~\ref{sec_3} we present the improved numerical scheme. In Sec.~\ref{sec_4} we conduct numerical experiments to demonstrate the validity of the scheme. Finally, we draw conclusions in Sec.~\ref{sec_5}.

\section{Grid and notations}
\label{sec_2}
\begin{figure*}
  \begin{center}
    \includegraphics[width=0.9\linewidth]{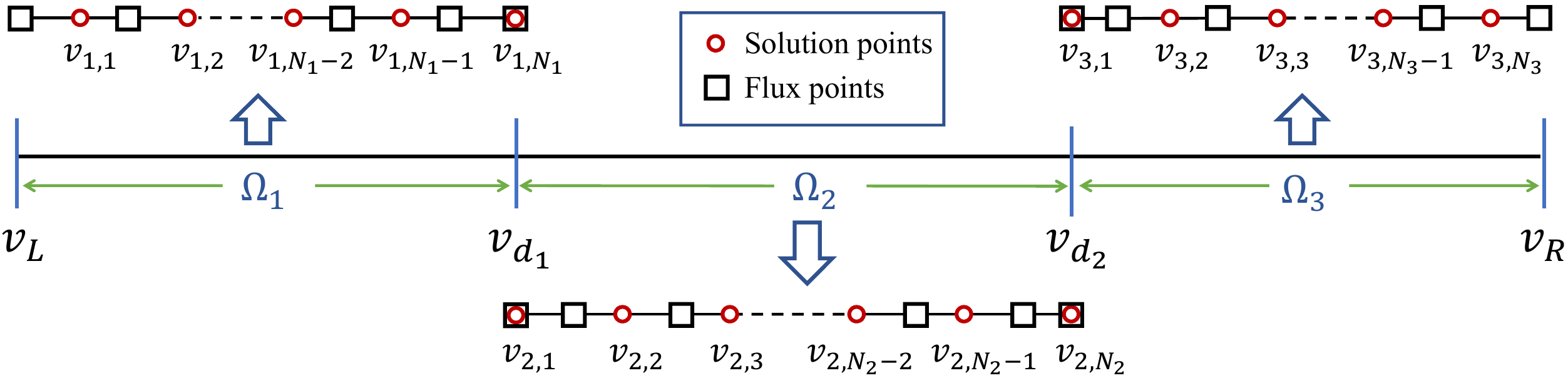}
  \end{center}
  \caption{The grid staggered by flux points and solution points for the case with two jumps at $v=v_{d_1}$ and $v_{d_2}$, respectively. The two jumps are set as both flux points and solution points and are used to divide the computational domain $[v_L, v_R]$ into three subdomains
  $\Omega_i$ ($i=1,2,3$). }\label{fig_1}
\end{figure*}

As Ref.~\cite{ChenDeng2018PRE}, we still describe the algorithm by assuming that there are two jumps in the drift. For the assumed computational domain
$[v_{_L},v_{_R}]$, we use the positions of these two jumps  (denoted by $v_{d_1}$ and $v_{d_2}$ with $v_{d_1}<v_{d_2}$) to divide it into three parts: $\Omega_1=[v_{_L}, v_{d_1}]$, $\Omega_2=[v_{d_1}, v_{d_2}]$ and $\Omega_3=[v_{d_2},v_{_R}]$; see Fig.~\ref{fig_1}.

The grid adopted here is almost the same as that in Ref.~\cite{ChenDeng2018PRE} (see Fig.~1 therein). The only difference is that the jumps are set as both solution points and flux points here, rather than only solution points in Ref.~\cite{ChenDeng2018PRE}. We will see later that this setup is very important for us to design an improved scheme.

In each subdomain $\Omega_i$ $(i=1,2,3)$, we have two sets of grid points:
\begin{align}
  \Omega_1:\; &\mathbf{v}_1=[v_{1,1},v_{1,2},\dots,v_{1,N_1}]^T,\nonumber \\
                 &\hat{\mathbf{v}}_1=[v_{1,1/2},v_{1,3/2},\dots,v_{1,N_1-1/2},v_{1,N_1}]^T, \label{ae}
\end{align}
\begin{align}
  \Omega_2:\; &\mathbf{v}_2=[v_{2,1},v_{2,2},\dots,v_{2,N_2}]^T,\nonumber \\
                 &\hat{\mathbf{v}}_2=[v_{2,1},v_{2,3/2},v_{2,5/2},\dots,v_{2,N_2-1/2},v_{2,N_2}]^T,\label{af}\\
  \Omega_3:\; &\mathbf{v}_3=[v_{3,1},v_{3,2},\dots,v_{3,N_3}]^T,\nonumber \\
                 &\hat{\mathbf{v}}_3=[v_{3,1},v_{3,3/2},v_{3,5/2},\dots,v_{3,N_3+1/2}]^T, \label{ag}
\end{align}
where $\mathbf{v}_i$ and $\hat{\mathbf{v}}_i$ denote the vectors of solution points and flux points (see Fig.~\ref{fig_1}), respectively. Here $N_i$ are the numbers of solution points in $\Omega_i$,
\begin{align}
  v_{1,j}=&v_{_L}+(j-1/2)h_1,&&v_{1,j+1/2}=v_{_L}+jh_1, \nonumber\\
  v_{2,j}=&v_{d_1}+(j-1)h_2, &&v_{2,j+1/2}=v_{d_1}+(j-1/2)h_2,\nonumber\\
  v_{3,j}=&v_{d_2}+(j-1)h_3, &&v_{3,j+1/2}=v_{d_2}+(j-1/2)h_3\nonumber
\end{align}
with $h_i$ denoting the spatial steps for the subdomains, defined by
\begin{align}
   \begin{cases}
   h_1=(v_{d_1}-v_{_L})/(N_1-1/2),\\
   h_2=(v_{d_2}-v_{d_1})/(N_2-1),\\
   h_3=(v_{_R}-v_{d_2})/(N_3-1/2).
   \end{cases}
   \label{ah}
\end{align}
Especially, we have $
   v_{1,N_1}=v_{2,1}=v_{d_1}
$ and
$
   v_{2,N_2}=v_{3,1}=v_{d_2}
$, showing that the discontinuous points are both solution points and
flux points; see Eqs.~(\ref{ae})-(\ref{ag}).

\section{Modified scheme}
\label{sec_3}

In this section, we consider how to compute the derivative $\partial_v f$. For convenience, let us denote the grid values corresponding to $\mathbf{v}_i$ by a column vector $\mathbf{p}_i$, which are known at a certain time level. 
Then similarly with the procedure of Ref.~\cite{ChenDeng2018PRE}, we first compute the term $\Phi(v)p$ in each subdomain as
\begin{equation}
  \max\{\Phi(\hat{v}_{i,j}),0\}\hat{p}_{i,j}^-+\min\{\Phi(\hat{v}_{i,j}),0\}\hat{p}_{i,j}^+,
  \label{ba}
\end{equation}
where $\hat{v}_{i,j}$ represents the $j$-th entry of the vector $\hat{\mathbf{v}}_{i}$, with the corresponding left and right values denoting by $\hat{p}_{i,j}^-$ and $\hat{p}_{i,j}^+$, respectively. In this work we compute the values of $\hat{p}_{i,j}^\pm$ by using the following interpolation procedure:
\begin{align}
   \Omega_1:\quad &\hat{\mathbf{p}}_{1}^+=I_1^+ \mathbf{p}_1^R, 
   && \hat{\mathbf{p}}_{1}^-=I_1^- [\hat{p}^+_{1,1},\mathbf{p}_1^T]^T, \label{bb_1}\\
   \Omega_2:\quad & \hat{\mathbf{p}}_{2}^+ =I_2^+ \mathbf{p}_2^R, && 
   \hat{\mathbf{p}}_{2}^- =I_2^- \mathbf{p}_2^L, \label{bb_2}\\
   \Omega_3:\quad & \hat{\mathbf{p}}_{3}^- =I_3^- \mathbf{p}_3^L, &&
   \hat{\mathbf{p}}_{3}^+ =I_3^+ [\mathbf{p}_3^T,\hat{p}^-_{3,N_3+1}]^T, \label{bb_3}
\end{align}
where $\mathbf{p}_1^R$ is the vector $\mathbf{p}_1$ with the last entry $p_{1,N_1}$ replacing by the average value $( p_{1,N_1} + p_{2,1} )/2$, $\mathbf{p}_2^R$ represents the vector $\mathbf{p}_2$ with the entry $p_{2,N_2}$ replacing by $( p_{2,N_2} + p_{3,1} )/2$, $\mathbf{p}_2^L$  stands for the vector $\mathbf{p}_2$ with the entry $p_{2,1}$ replacing by $( p_{1,N_1} + p_{2,1} )/2$, and $\mathbf{p}_3^L$ denotes the vector $\mathbf{p}_3$ with the entry $p_{3,1}$ replacing by $( p_{2,N_2} + p_{3,1} )/2$. For convenience of the presentation, the corresponding fifth-order interpolation matrices $I_i^\pm$ are presented in Appendix \ref{sec_appa1}.
It is noted that the derivation of these matrices is exactly the same as that in Ref.~\cite{ChenDeng2018PRE}. Therefore, we omit the details here and present the expressions explicitly only in the Appendix. This claim applies for the rest of this section as well.

 Second, we compute the derivative $\partial_v p$ at the points $\hat{v}_{i,j}$ by using difference schemes in each subdomain,
  \begin{align}
   \hat{\mathbf{p}}_{v,1}&=\frac{1}{h_1}A_1\mathbf{p}_1^R,\label{bc_1}\\
   \hat{\mathbf{p}}_{v,2}&=\frac{1}{h_2}A_2\mathbf{p}_2^{LR},\label{bc_2}\\
   \hat{\mathbf{p}}_{v,3}&=\frac{1}{h_3}A_3\mathbf{p}_3^L,\label{bc_3}
 \end{align}
where $\hat{\mathbf{p}}_{v,i}$ are the column vectors with the entries $\hat{p}_{v,i,j}$ corresponding to $\hat{v}_{i,j}$, and $A_i$ are difference matrices with expressions presented in Appendix \ref{sec_appa2}. Here
$\mathbf{p}_2^{LR}$ denotes the vector $\mathbf{p}_2^{R}$ (\ref{bb_2}) with the first entry replacing by $(p_{1,N_1}+p_{2,1})/2$.
Then from Eqs.~(\ref{ad}) and (\ref{ba})  we obtain immediately the values of the flux corresponding to $\hat{v}_{i,j}$,
 \begin{align}
   \hat{f}_{i,j}=& -\max\{\Phi(\hat{v}_{i,j}),0\}\hat{p}_{i,j}^-\nonumber\\
  & -\min\{\Phi(\hat{v}_{i,j}),0\}\hat{p}_{i,j}^+
  +D\hat{p}_{v,i,j}.
  \label{bd}
 \end{align}
At this step, we shall impose the continuous condition of the flux by store
the average values
 \begin{equation}
   \bar{f}_{1}=\frac{1}{2}(\hat{f}_{1,N_1+1}+\hat{f}_{2,1}), \quad \bar{f}_{2}=\frac{1}{2}(\hat{f}_{2,N_2+1}+\hat{f}_{3,1}),
   \label{bu}
 \end{equation}
and modify the values of the flux at the jumps of the drift as
 \begin{equation}
   \hat{f}_{1,N_1+1}=\hat{f}_{2,1}=\bar{f}_{1}, \quad \hat{f}_{2,N_2+1}=\hat{f}_{3,1}=\bar{f}_{2}.
   \label{bv}
 \end{equation}

 Thirdly, rather than deriving a difference scheme to compute the derivative $\partial_vf$ in the entire computational domain as Ref.~\cite{ChenDeng2018PRE}, we still derive difference schemes in each subdomain to avoid losing accuracy. In each subdomain $\Omega_i$, the scheme is denoted by
  \begin{equation}
   \mathbf{f}_{v,i}=\frac{1}{h_i}D_i \hat{\mathbf{f}}_i,
   \label{be}
 \end{equation}
 where the difference matrices $D_i$ are presented in Appendix \ref{sec_appa3}.

 Finally, we have the ordinary differential equations
 \begin{equation}
   \frac{d}{dt}\mathbf{p}_{i}=\mathbf{f}_{v,i},
   \label{bf}
 \end{equation}
 which can be solved by some time-marching schemes.
 Here the third-order Runge-Kutta scheme as used in Ref.~\cite{ChenDeng2018PRE}
 is still employed [see Eqs.~(30) and (31) therein]. It should be mentioned that at each time step we shall update the values as
\begin{equation}
  \mathbf{p}_1=\mathbf{p}_1^R, \quad
  \mathbf{p}_2=\mathbf{p}_2^{LR},\quad
  \mathbf{p}_3=\mathbf{p}_3^L
\end{equation}
due to the continuous condition of the propagator. Here $\mathbf{p}_1^R$, $\mathbf{p}_2^{LR}$ and $\mathbf{p}_3^L$ are defined in Eqs.~(\ref{bb_1}),
(\ref{bc_2}) and (\ref{bb_3}), respectively. 

\section{Numerical experiments}
\label{sec_4}

In this section, we present two numerical examples to validate the modified scheme. For the first example, we choose the exact solution $p(v,\tau_0|v_0,0)$ to be the initial condition and start the computation from time $t=\tau_0$.
For the second example, while the exact solution is not available, we just set the initial condition to be Gaussian,
\begin{equation}
   p(v,\tau_0|v_0,0)=\frac{1}{ \sqrt{4\pi D \tau_0} }e^{ -[v-v_0-\Phi(v_0)\tau_0]^2/(4D\tau_0) }.
   \label{ca}
\end{equation}
For convenience, $D=0.5$ and $\tau_0=0.01$ are chosen for both cases. The $L^2$ error and $L^\infty$ error, presented in this section \red{for the case with one jump (see Tab.~\ref{tab_2}), 
are defined by
\begin{gather*}
  L^2 \mbox{ error}=\sqrt{\sum_{j=1}^{N_1-1} e_{1,j}^2h_1+\frac{1}{2}e^2_{1,N_1}(h_1+h_2)+\sum_{j=2}^{N_2}e_{2,j}^2h_2},\\
  L^\infty \mbox{ error}=\max_{i,j}\{|e_{i,j}|\},
\end{gather*}
where $e_{i,j}=p_{i,j}-p(v_{i,j})$ are the errors between numerical results and exact solutions.
For other cases, the errors are defined similarly.}

\subsection{Pure dry friction}
We first consider the case with the drift $\Phi(v)=-\mu\,\mbox{sgn}(v)$, where
$\mu>0$ and $\mbox{sgn}(v)$ denotes the sign of $v$. In this case, Eq.~(\ref{aa}) represents the Brownian motion with pure dry friction \cite{Gennes2005}, whose propagator can be obtained analytically \cite{CaugheyDienes1961,Karatzas1984,TouchetteStraetenJust2010} as
\begin{align}
    p(v,t|v_0,0)=\tfrac{\mu}{D}\hat{p}\left( \tfrac{\mu}{D}v,\tfrac{\mu^2}{D}t\big|\tfrac{\mu}{D}v_0,0 \right),
    \label{cc}
\end{align}
where
\begin{align*}
  \hat{p}(x,\tau|x_0,0)=& \frac{e^{-\tau/4}}{2\sqrt{\pi \tau}}e^{-(|x|-|x_0|)/2}e^{-(x-x_0)^2/(4\tau)} \nonumber\\
  &  + \frac{e^{-|x|}}{4}\left[
    1+\mathrm{erf}\left( \tfrac{\tau-(|x|+|x_0|)}{2\sqrt{\tau}} \right)
  \right]
\end{align*}
with $ \mathrm{erf}(x)=2\int_0^x \exp(-z^2)dz/\sqrt{\pi}
$ denoting the error function.

Here we divide the computational domain $[-8,8]$ into two subdomains by using the point $v_d=0$, and set the time step to be $\tau=0.4\min\{h_1^2,h_2^2\}$.
In addition, we choose $\mu=1$, $v_0=2$ and set a zero current condition for both the left and the right
boundaries. As we can see in Tab.~\ref{tab_2}, the numerical results show that
the scheme achieves the claimed fifth-order convergence rate for this case
with a discontinuity in the drift. It is evident that this modified scheme improves
the accuracy significantly compared with the scheme presented in Ref.~\cite{ChenDeng2018PRE} (see Tab.~IV therein).

\begin{table}
   \begin{tabular*}{\linewidth}{@{\extracolsep{\fill}}ccccccc}
      \hline\hline
      $N_1$ &$N_2$ & $N_v$ & $L^2$ error & Rate & $L^\infty$ error & Rate \\
      \hline
   50 & 50  & 99 & 2.86E-04 &--& 2.23E-04 &--
\\ 100 & 100  & 199 & 3.07E-06 &6.50 & 6.91E-06	&4.98 
\\ 200 & 200 & 399 & 9.78E-08 &	4.95 & 2.50E-07 &4.77 
\\ 400 & 400 & 799 & 3.04E-09 & 5.00 & 8.38E-09 &4.89 
\\ 800 & 800 & 1599 & 9.45E-11& 5.00 & 2.71E-10 &4.94 
\\\hline\hline
   \end{tabular*}
   \caption{Accuracy test for Eq.~(\ref{ac}) with $\Phi(v)=-\mathrm{sgn}(v)$ at time $t=1$. $v_0=2$ and $N_v=N_1+N_2-1$. }\label{tab_2}
\end{table}

\subsection{Drift with two discontinuities}
\begin{figure}
  \begin{center}
    \includegraphics[width=0.8\linewidth]{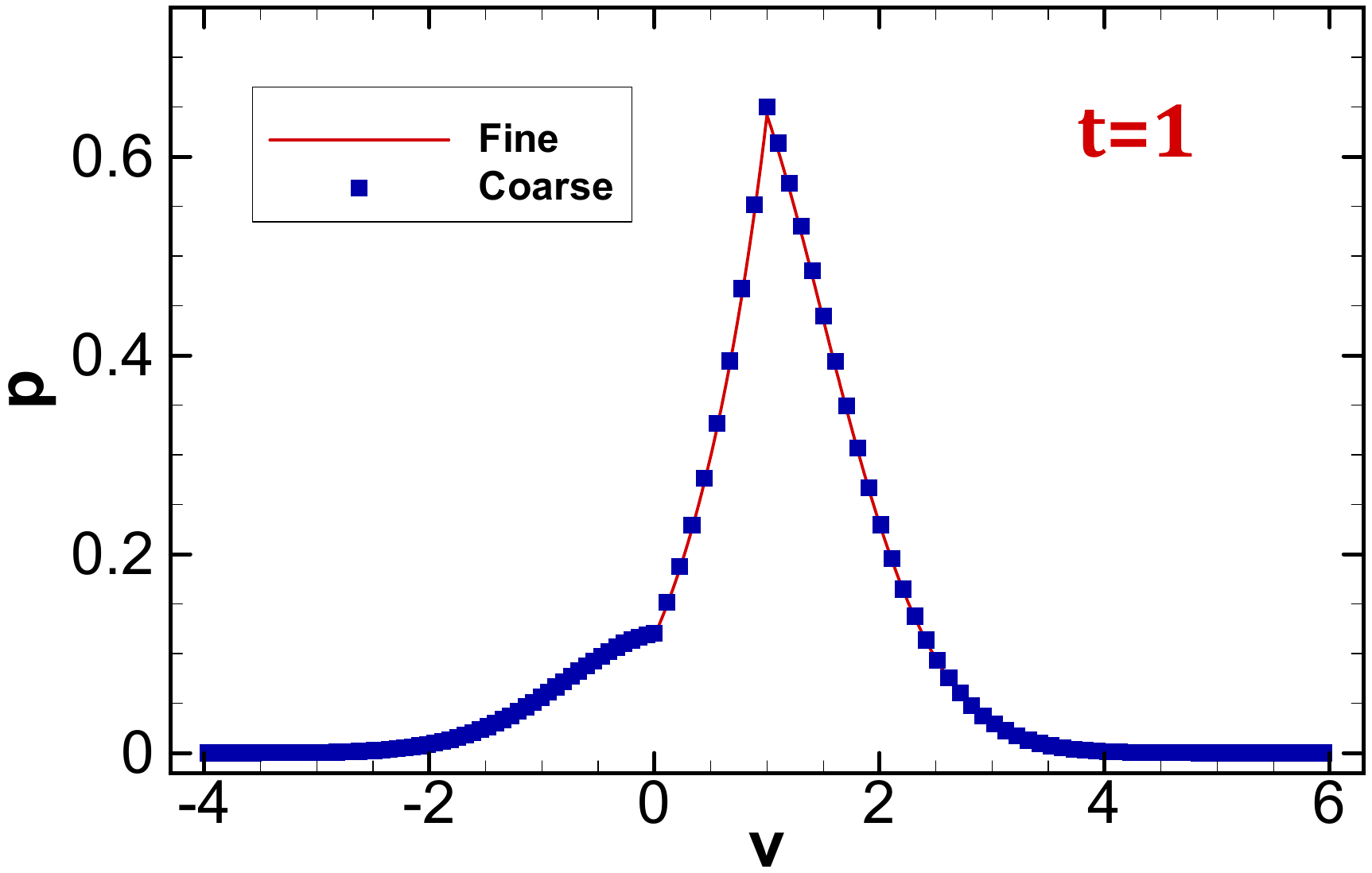}
  \end{center}
  \caption{Propagator of the case with the drift (\ref{cc1}) at time $t=1$; $v_0=0.5$. $N_1=60$, $N_2=10$ and $N_3 = 50$ are chosen for the coarse grid, while $N_1=480$, $N_2=80$ and 
$N_3 = 400$ are set for the fine grid.}
  \label{fig_2}
\end{figure}

To show that the modified scheme also works for the case with more discontinuities in the drift, we take as an example the case with two discontinuities. To be specific, we consider the following drift
\cite{Dereudre2017}:
\begin{align}
   \Phi(v)=\begin{cases}
     0, & v<0,\\
     1, & 0<v<1,\\
     0, & v>1.
   \end{cases}
   \label{cc1}
\end{align}
The computational domain is set to be $[-4,6]$, divided into three subdomains by the discontinuous points $v_{d_1}=0$ and $v_{d_2}=1$. The time step $\tau=0.4\min\{h_1^2,h_2^2,h_3^2\}$ is applied for implementing the time scheme. In this case, we do not have to specify boundary value condtions for the proposed scheme.  As we can see in Fig.~\ref{fig_2}, the coarse-grid solution matches well with the fine-grid solution. The test of accuracy is not considered here since the exact solution of the case is not available. But we can observe that the obtained solution profiles agree with the result presented in Refs.~\cite{Dereudre2017} and  \cite{ChenDeng2018PRE} [see Figs.~6(a) and 8(a) therein, respectively].   

\section{Conclusions}
\label{sec_5}

In this paper, we have proposed a modified scheme to improve the scheme presented in Ref.~\cite{ChenDeng2018PRE}. The idea is quite simple, just avoiding using interpolation schemes or difference schemes across discontinuities of the drift. The grid with discontinuities setting to be both solution points and flux points is the key for using the two matching conditions for each discontinuity. At each jump of the drift, a simple average procedure is implemented to incorporate the matching conditions with respect to the propagator and the flux. To demonstrate that the proposed scheme indeed improves the result from second order to fifth order, we have presented the result for the case with pure dry friction. In addition, the case with two jumps have also been computed to validate the scheme. 

\red{Although we have only considered here the simple one-dimensional case, the proposed scheme can certainly be applied to high-dimensional problems  with drift-admitting jumps \cite{DasPuri2017}, simply in a dimension-by-dimension manner. Moreover, we remark that the scheme is also expected to work for other cases with (additive or multiplicative) colored noises \cite{GeffertJust2017}. 
}

\begin{acknowledgments}
  This work was supported by the National Natural Science Foundation of China (Grant No. 11601517).
\end{acknowledgments}


%

\appendix
\section{Coefficient matrices for interpolation schemes and difference schemes}
\label{sec_appa}

For convenience of the reader, the coefficient matrices of the schemes (\ref{bb_1})-(\ref{bb_3}), (\ref{bc_1})-(\ref{bc_3}) and (\ref{be}) are presented directly here. 

\subsection{Coefficient matrices for Eqs.~(\ref{bb_1})-(\ref{bb_3})}\label{sec_appa1}

It is noted that the size of the matrices $I_i^\pm$ can be easily observed from Eqs.~(\ref{bb_1})-(\ref{bb_3}). Here we just present the values as follows: 
\begin{widetext}
\begin{equation*}
I_1^+=\begin{bmatrix}
 \frac{315}{128} & -\frac{105}{32} & \frac{189}{64} &
   -\frac{45}{32} & \frac{35}{128} &  &  \\[3pt]
 \frac{35}{128} & \frac{35}{32} & -\frac{35}{64} & \frac{7}{32} &
   -\frac{5}{128} &  &  \\[3pt]
 -\frac{5}{128} & \frac{15}{32} & \frac{45}{64} & -\frac{5}{32} &
   \frac{3}{128} &  &  \\[3pt]
 & \ddots & \ddots & \ddots & \ddots &
   \ddots \\[3pt]
 &  & -\frac{5}{128} & \frac{15}{32} & \frac{45}{64} &
   -\frac{5}{32} & \frac{3}{128} \\[3pt]
 &  & \frac{3}{128} & -\frac{5}{32} & \frac{45}{64} &
   \frac{15}{32} & -\frac{5}{128} \\[3pt]
 &  & -\frac{5}{128} & \frac{7}{32} & -\frac{35}{64} &
   \frac{35}{32} & \frac{35}{128}\\[3pt]
 &  &  &  &  &  &  1  
\end{bmatrix},\quad 
I_1^-=\begin{bmatrix}
 1 & \\[3pt]
 -\frac{1}{7} & \frac{5}{8} & \frac{5}{8} & -\frac{1}{8} &
   \frac{1}{56} \\[3pt]
 \frac{3}{35} & -\frac{1}{4} & \frac{3}{4} & \frac{9}{20} &
   -\frac{1}{28} \\[3pt]
 & \frac{3}{128} & -\frac{5}{32} & \frac{45}{64} & \frac{15}{32}
 & -\frac{5}{128} \\[3pt]
 & & \ddots & \ddots & \ddots & \ddots &
   \ddots \\[3pt]
 & & & \frac{3}{128} & -\frac{5}{32} & \frac{45}{64} &
   \frac{15}{32} & -\frac{5}{128} \\[3pt]
 & & & -\frac{5}{128} & \frac{7}{32} & -\frac{35}{64} &
   \frac{35}{32} & \frac{35}{128}\\[3pt]
 &  &  &  &  &  &  & 1  
\end{bmatrix},
\end{equation*}
\begin{equation*}
    I_2^+=\begin{bmatrix}
    1\\[3pt]
   \frac{35}{128} & \frac{35}{32} & -\frac{35}{64} & \frac{7}{32} &
   -\frac{5}{128} \\[3pt]
 -\frac{5}{128} & \frac{15}{32} & \frac{45}{64} & -\frac{5}{32} &
   \frac{3}{128} \\[3pt]
 & \ddots & \ddots & \ddots & \ddots &
   \ddots  \\[3pt]
 & & -\frac{5}{128} & \frac{15}{32} & \frac{45}{64} &
   -\frac{5}{32} & \frac{3}{128} \\[3pt]
 & & \frac{3}{128} & -\frac{5}{32} & \frac{45}{64} &
   \frac{15}{32} & -\frac{5}{128} \\[3pt]
 & & -\frac{5}{128} & \frac{7}{32} & -\frac{35}{64} &
   \frac{35}{32} & \frac{35}{128} \\[3pt] 
   &  &  &  &  &  &  1  
  \end{bmatrix},
\quad
  I_2^-=\begin{bmatrix}
  1\\[3pt]
   \frac{35}{128} & \frac{35}{32} & -\frac{35}{64} & \frac{7}{32} &
   -\frac{5}{128} \\[3pt]
 -\frac{5}{128} & \frac{15}{32} & \frac{45}{64} & -\frac{5}{32} &
   \frac{3}{128}  \\[3pt]
 \frac{3}{128} & -\frac{5}{32} & \frac{45}{64} & \frac{15}{32} &
   -\frac{5}{128}  \\[3pt]
 & \ddots & \ddots & \ddots & \ddots &
   \ddots  \\[3pt]
 & & \frac{3}{128} & -\frac{5}{32} & \frac{45}{64} &
   \frac{15}{32} & -\frac{5}{128} \\[3pt]
 & & -\frac{5}{128} & \frac{7}{32} & -\frac{35}{64} &
   \frac{35}{32} & \frac{35}{128}\\[3pt]
   &  &  &  &  &   & 1  
  \end{bmatrix},
  \end{equation*}
\begin{equation*}
    I_3^+=\begin{bmatrix}
    1\\[3pt]
\frac{35}{128} & \frac{35}{32} & -\frac{35}{64} & \frac{7}{32} &
   -\frac{5}{128}  \\[3pt]
 -\frac{5}{128} & \frac{15}{32} & \frac{45}{64} & -\frac{5}{32} &
   \frac{3}{128} \\[3pt]
 & \ddots & \ddots & \ddots & \ddots &
   \ddots  \\[3pt]
 & & -\frac{5}{128} & \frac{15}{32} & \frac{45}{64} &
   -\frac{5}{32} & \frac{3}{128}  \\[3pt]
 & & & -\frac{1}{28} & \frac{9}{20} & \frac{3}{4} &
   -\frac{1}{4} & \frac{3}{35} \\[3pt]
 & & & \frac{1}{56} & -\frac{1}{8} & \frac{5}{8} &
   \frac{5}{8} & -\frac{1}{7} \\[3pt]
  & & & & & & &1
  \end{bmatrix},
  \quad
  I_3^-=\begin{bmatrix}
  1\\[3pt]
  \frac{35}{128} & \frac{35}{32} & -\frac{35}{64} & \frac{7}{32} &
   -\frac{5}{128}  \\[3pt]
 -\frac{5}{128} & \frac{15}{32} & \frac{45}{64} & -\frac{5}{32} &
   \frac{3}{128}  \\[3pt]
 \frac{3}{128} & -\frac{5}{32} & \frac{45}{64} & \frac{15}{32} &
   -\frac{5}{128}  \\[3pt]
 & \ddots & \ddots & \ddots & \ddots &
   \ddots  \\[3pt]
 & & \frac{3}{128} & -\frac{5}{32} & \frac{45}{64} &
   \frac{15}{32} & -\frac{5}{128} \\[3pt]
 & & -\frac{5}{128} & \frac{7}{32} & -\frac{35}{64} &
   \frac{35}{32} & \frac{35}{128} \\[3pt]
 & & \frac{35}{128} & -\frac{45}{32} & \frac{189}{64} &
   -\frac{105}{32} & \frac{315}{128}
  \end{bmatrix}.
  \end{equation*}
\end{widetext}
\subsection{Coefficient matrices for Eqs.~(\ref{bc_1})-(\ref{bc_3})}\label{sec_appa2}

In Eqs.~(\ref{bc_1})-(\ref{bc_3}), the difference matrices $A_i$ with sizes of 
$(N_i+1)\times N_i$ read as follows:
\begin{equation*}
  A_1=\begin{bmatrix}
    -\frac{31}{8} & \frac{229}{24} & -\frac{75}{8} & \frac{37}{8} &
   -\frac{11}{12} \\[3pt]
 -\frac{11}{12} & \frac{17}{24} & \frac{3}{8} & -\frac{5}{24} &
   \frac{1}{24}  \\[3pt]
 \frac{1}{24} & -\frac{9}{8} & \frac{9}{8} & -\frac{1}{24} \\[3pt]
 -\frac{3}{640} & \frac{25}{384} & -\frac{75}{64} & \frac{75}{64}
   & -\frac{25}{384} & \frac{3}{640} \\[3pt]
  & \ddots & \ddots & \ddots & \ddots &
   \ddots &
   \ddots \\[3pt]
 & & -\frac{3}{640} & \frac{25}{384} & -\frac{75}{64} &
   \frac{75}{64} & -\frac{25}{384} & \frac{3}{640} \\[3pt]
 & & & & \frac{1}{24} & -\frac{9}{8} & \frac{9}{8} &
   -\frac{1}{24} \\[3pt]
 & & & -\frac{1}{24} & \frac{5}{24} & -\frac{3}{8} &
   -\frac{17}{24} & \frac{11}{12} \\[3pt]
 & & & \frac{1}{4} & -\frac{4}{3} & 3 & -4 & \frac{25}{12}
  \end{bmatrix},
\end{equation*}

\begin{equation*}
  A_2=\begin{bmatrix}
   -\frac{25}{12} & 4 & -3 & \frac{4}{3} & -\frac{1}{4} \\[3pt]
 -\frac{11}{12} & \frac{17}{24} & \frac{3}{8} & -\frac{5}{24} &
   \frac{1}{24} \\[3pt]
 \frac{1}{24} & -\frac{9}{8} & \frac{9}{8} & -\frac{1}{24} \\[3pt]
 -\frac{3}{640} & \frac{25}{384} & -\frac{75}{64} & \frac{75}{64}
   & -\frac{25}{384} & \frac{3}{640} \\[3pt]
 & \ddots & \ddots & \ddots & \ddots &
   \ddots &
   \ddots  \\[3pt]
 & & -\frac{3}{640} & \frac{25}{384} & -\frac{75}{64} &
   \frac{75}{64} & -\frac{25}{384} & \frac{3}{640} \\[3pt]
 & & & & \frac{1}{24} & -\frac{9}{8} & \frac{9}{8} &
   -\frac{1}{24} \\[3pt]
 & & & -\frac{1}{24} & \frac{5}{24} & -\frac{3}{8} &
   -\frac{17}{24} & \frac{11}{12} \\[3pt]
 & & & \frac{1}{4} & -\frac{4}{3} & 3 & -4 & \frac{25}{12}
  \end{bmatrix},
\end{equation*}

\begin{equation*}
  A_3=\begin{bmatrix}
  -\frac{25}{12} & 4 & -3 & \frac{4}{3} & -\frac{1}{4}
   \\[3pt]
 -\frac{11}{12} & \frac{17}{24} & \frac{3}{8} & -\frac{5}{24} &
   \frac{1}{24}  \\[3pt]
 \frac{1}{24} & -\frac{9}{8} & \frac{9}{8} & -\frac{1}{24}  \\[3pt]
 -\frac{3}{640} & \frac{25}{384} & -\frac{75}{64} & \frac{75}{64}
   & -\frac{25}{384} & \frac{3}{640}  \\[3pt]
 & \ddots & \ddots & \ddots & \ddots &
   \ddots&
   \ddots  \\[3pt]
 & & -\frac{3}{640} & \frac{25}{384} & -\frac{75}{64} &
   \frac{75}{64} & -\frac{25}{384} & \frac{3}{640} \\[3pt]
 & & & & \frac{1}{24} & -\frac{9}{8} & \frac{9}{8} &
   -\frac{1}{24} \\[3pt]
 & & & -\frac{1}{24} & \frac{5}{24} & -\frac{3}{8} &
   -\frac{17}{24} & \frac{11}{12} \\[3pt]
 & & & \frac{11}{12} & -\frac{37}{8} & \frac{75}{8} &
   -\frac{229}{24} & \frac{31}{8}
  \end{bmatrix}.
\end{equation*}

\subsection{Coefficient matrices for Eq.~(\ref{be})}\label{sec_appa3}

In Eq.~(\ref{be}), the difference matrices $D_i$ with sizes of 
$N_i\times (N_i+1)$ read as follows:
\begin{equation*}
    D_1=\begin{bmatrix}
-\frac{11}{12} & \frac{17}{24} & \frac{3}{8} & -\frac{5}{24} &
   \frac{1}{24} \\[3pt]
 \frac{1}{24} & -\frac{9}{8} & \frac{9}{8} & -\frac{1}{24} \\[3pt]
 -\frac{3}{640} & \frac{25}{384} & -\frac{75}{64} & \frac{75}{64}
   & -\frac{25}{384} & \frac{3}{640} \\[3pt]
 & \ddots & \ddots & \ddots & \ddots &
   \ddots &
   \ddots \\[3pt]
 & & -\frac{3}{640} & \frac{25}{384} & -\frac{75}{64} &
   \frac{75}{64} & -\frac{25}{384} & \frac{3}{640} \\[3pt]
 & & & & \frac{1}{24} & -\frac{9}{8} & \frac{9}{8} &
   -\frac{1}{24} \\[3pt]
 & & & & -\frac{1}{168} & \frac{3}{40} & -\frac{29}{24} &
   \frac{31}{24} & -\frac{16}{105} \\[3pt]
 & & & & \frac{5}{56} & -\frac{21}{40} & \frac{35}{24} &
   -\frac{35}{8} & \frac{352}{105}
  \end{bmatrix},
\end{equation*}
\begin{widetext}
\begin{equation*}
    D_2=\begin{bmatrix}
   -\frac{352}{105} & \frac{35}{8} & -\frac{35}{24} & \frac{21}{40}
   & -\frac{5}{56}  \\[3pt]
 \frac{16}{105} & -\frac{31}{24} & \frac{29}{24} & -\frac{3}{40} &
   \frac{1}{168}  \\[3pt]
  & \frac{1}{24} & -\frac{9}{8} & \frac{9}{8} & -\frac{1}{24} \\[3pt]
 & -\frac{3}{640} & \frac{25}{384} & -\frac{75}{64} &
   \frac{75}{64} & -\frac{25}{384} & \frac{3}{640} \\[3pt]
 & & \ddots & \ddots & \ddots & \ddots &
   \ddots &
   \ddots \\[3pt]
  &  &  & -\frac{3}{640} & \frac{25}{384} & -\frac{75}{64} &
   \frac{75}{64} & -\frac{25}{384} & \frac{3}{640}  \\[3pt]
  &  &  &  &  & \frac{1}{24} & -\frac{9}{8} & \frac{9}{8} &
   -\frac{1}{24} &  \\[3pt]
 & & & & & -\frac{1}{168} & \frac{3}{40} &
   -\frac{29}{24} & \frac{31}{24} & -\frac{16}{105} \\[3pt]
  &  &  &  &  & \frac{5}{56} & -\frac{21}{40} & \frac{35}{24}
   & -\frac{35}{8} & \frac{352}{105}
  \end{bmatrix},
\end{equation*}
\begin{equation*}
   D_3=\begin{bmatrix}
  -\frac{352}{105} & \frac{35}{8} & -\frac{35}{24} & \frac{21}{40}
   & -\frac{5}{56}  \\[3pt]
 \frac{16}{105} & -\frac{31}{24} & \frac{29}{24} & -\frac{3}{40} &
   \frac{1}{168}  \\[3pt]
 & \frac{1}{24} & -\frac{9}{8} & \frac{9}{8} & -\frac{1}{24} \\[3pt]
 & -\frac{3}{640} & \frac{25}{384} & -\frac{75}{64} &
   \frac{75}{64} & -\frac{25}{384} & \frac{3}{640}  \\[3pt]
 & & \ddots & \ddots & \ddots & \ddots &
   \ddots&
   \ddots \\[3pt]
 & & & -\frac{3}{640} & \frac{25}{384} & -\frac{75}{64} &
   \frac{75}{64} & -\frac{25}{384} & \frac{3}{640} \\[3pt]
 & & & & & \frac{1}{24} & -\frac{9}{8} & \frac{9}{8} &
   -\frac{1}{24} \\[3pt]
 & & & & -\frac{1}{24} & \frac{5}{24} & -\frac{3}{8} &
   -\frac{17}{24} & \frac{11}{12}
  \end{bmatrix}.
\end{equation*}
\end{widetext}

\end{document}